\pdfoutput=1

\documentclass[preprint,journal]{vgtc}       

\usepackage{mathptmx}
\usepackage{graphicx}
\usepackage{times}
\usepackage{caption} 
\usepackage[bookmarks,backref=true,linkcolor=black]{hyperref} 
\hypersetup{
  pdfauthor = {},
  pdftitle = {},
  pdfsubject = {},
  pdfkeywords = {},
  colorlinks=true,
  linkcolor= black,
  citecolor= black,
  pageanchor=true,
  urlcolor = black,
  plainpages = false,
  linktocpage
}

\onlineid{0}

\title{A Novel Approach to Visualizing Dark Matter Simulations}

\author{Ralf Kaehler, Oliver Hahn, and Tom Abel}
\authorfooter{
\item
 Ralf Kaehler is with KIPAC/SLAC, E-mail: kaehler@slac.stanford.edu.
\item
 Oliver Hahn is with Stanford/SLAC, E-mail: ohahn@stanford.edu.
\item
  Tom Abel is with KIPAC/SLAC, E-mail:tabel@slac.stanford.edu.
}

\shortauthortitle{Biv \MakeLowercase{\textit{et al.}}: Visualizing Dark Matter Simulations}

\abstract{In the last decades cosmological N-body dark matter simulations have
  enabled ab initio studies of the formation of structure 
  in the Universe. Gravity amplified small density fluctuations
  generated shortly after the Big Bang, leading to the formation of galaxies
  in the cosmic web. These calculations have led to a growing demand
  for methods to analyze time-dependent particle based simulations.
  Rendering methods for such N-body simulation data
  usually employ some kind of splatting approach via point based
  rendering primitives and approximate the spatial distributions of
  physical quantities using kernel interpolation techniques,  common
  in SPH ({\sl Smoothed Particle Hydrodynamics})-codes.
  This paper proposes three GPU-assisted rendering
  approaches, based on a new, more accurate method to compute the physical
  densities of dark matter simulation data. It uses full 
  phase-space information to generate a tetrahedral tessellation of the
  computational domain, with mesh vertices defined by the simulation's
  dark matter particle positions. Over time the mesh is deformed by
  gravitational forces, causing the tetrahedral cells to warp and
  overlap. The new methods are well suited to visualize
  the cosmic web. In particular they preserve caustics,
  regions of high density that emerge, when several streams of dark matter particles
  share the same location in space, indicating the formation of
  structures like sheets, filaments and halos.
  We demonstrate the superior image quality of the new
  approaches in a comparison with three standard rendering techniques 
  for N-body simulation data.
} 


\teaser{
\centering
\includegraphics[width=16.35cm]{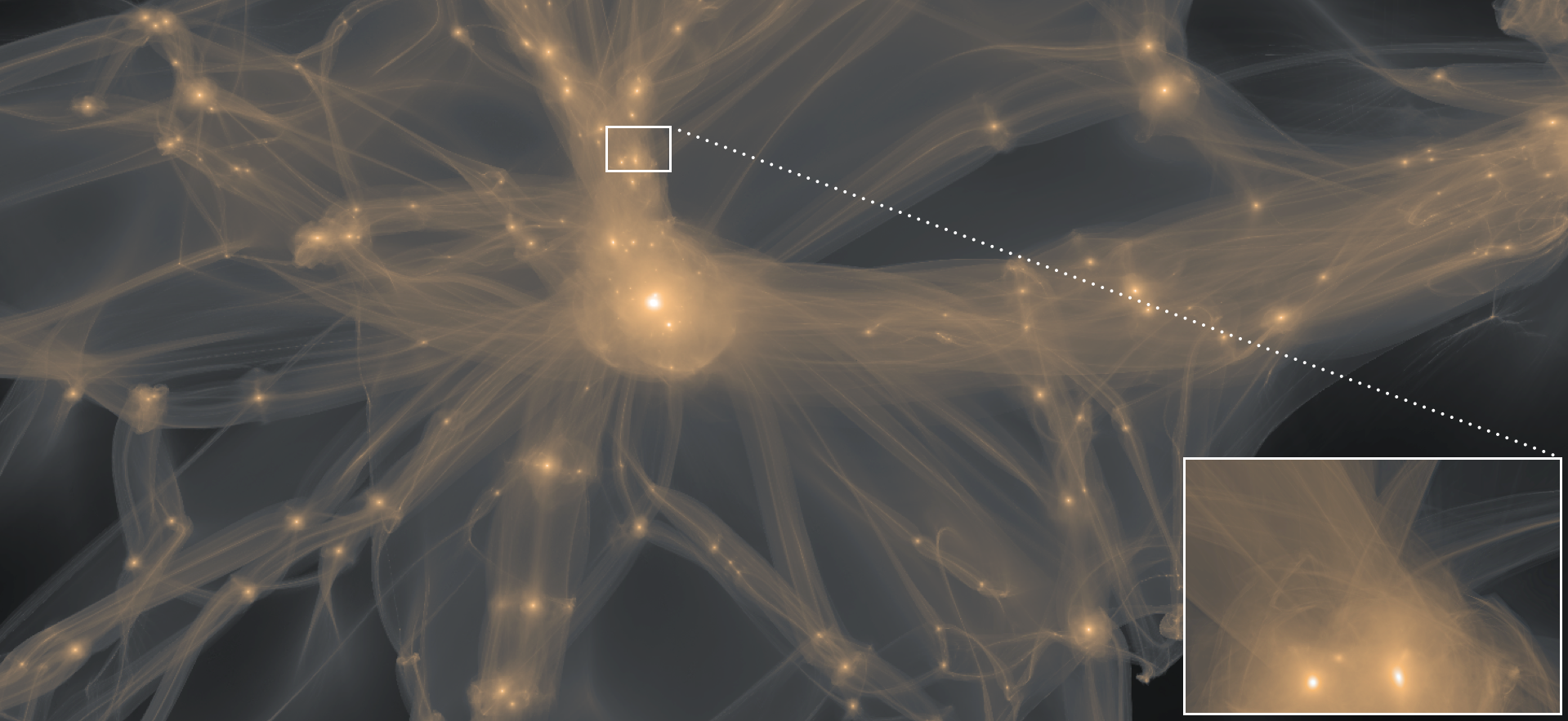}
\caption{\label{Fig-Teaser}%
   The density distribution rendered from a dark matter simulation
   using the tetrahedral tessellation approach applied in this paper.
   Large-scale structures like sheets (gray), filaments (yellow)
  and halos (white), as well as caustics on smaller scales (close-up), become clearly visible.}
}

\renewcommand{\manuscriptnotetxt}{}

\begin{document}

\firstsection{Introduction}

\maketitle


Starting with studies of the dynamics of clusters of galaxies by
Zwicky in the early 30's of the last century~\cite{1937ApJzwicky},
lots of observational evidence has been gathered, suggesting that the
luminous matter in the Universe, including objects like gas clouds and
stars, comprises only a tiny fraction of its total mass.  Most of the
mass in the Universe is thought to be cold dark matter.  ``Cold'' because
it moves at non--relativistic speeds and dark because it does not
interact with photons, and thus does not emit or absorb light, so that
its presence can only be measured through its gravitational influence
on ordinary matter.  
Some promising candidates to explain its nature are provided by
particle physics. The most popular is a light  neutralino suggested by
super--symmetric extensions of the standard model of particle
physics. 

Dark matter is the key ingredient in the formation of the large-scale
structure in the Universe,
which arise from small density fluctuations. These are
thought to have originated from quantum fluctuations and were stretched to macroscopic scales
during an early inflationary epoch shortly after the Big Bang.
Dark matter can then be thought of as a gas in which the
particles do not collide. To study its evolution, N-body simulations, originally developed in plasma physics
and for stellar dynamics, are being used~\cite{1985ApJS...57..241E,Springel05thecosmological}. 
The outcome of such simulations allow for
comparison with observational data of the large--scale distribution of
galaxies,  as for example the {\sl Sloan Digital Sky
  Survey}~\cite{2005ApJsloan}. Indeed, comparing such simulations with observational data 
dominates how the standard model of structure formation is being
tested. The simulation codes usually treat dark matter as a
collisionless gas sampled by a discrete number of tracer particles of equal
mass. These are propagated over time by the aggregated gravitational
forces acting on each particle. Different numerical methods
predominantly differ in how they compute the overall gravitational
forces in the computation domain.

Most previous visualizations of such simulations
projected each particle separately into screen space, using
different kernel profiles and methods to scale the splat sizes,
usually based on certain local interpolation schemes for the physical
quantities. One method that is particularly popular is based on
gathering the nearest n-neighbors for each particle and use adaptive kernel
smoothing to obtain a mass density at each particle position,
see e.g. Monaghan~\cite{1988CoPhC..48...89M}. This
approach necessarily introduces significant smoothing, especially in
regions of low dark matter densities, so--called {\sl voids}.

We present a rendering approach that is based on an
improved way to compute the dark matter density
distribution~\cite{2011AbelHahnKaehler,Shandarin:2011jv}. Instead of
operating solely on positional information, it uses the full phase-space information
available in N-body dark matter simulations. The computational domain
is tessellated using tetrahedral cells that contain equal amounts of
mass. The vertices of this mesh are defined by the dark matter
particles evolved by the N-body code. The connectivity of the cells is
generated once and is kept constant over time. Due to gravitational
forces each cell will be warped and at later times many cells will
overlap. However, the total mass per cell will stay constant, only its
density will change, due to the change of the spatial volume of the
cell. Accumulating the density contributions from all cells that fall
onto a certain location in the computational domain provides an
accurate density estimate for this region.
{
The contributions of this paper can be summarized as follows
\begin{itemize} 
\item  A data storage and access method that is tailored to the 
  specific properties of the underlying tessellation derived from the
  tracer particles of N-body dark matter simulations. It allows to generate the complete
  tessellation, including all connectivity information and derived
  quantities, like mass density per cell, on-the-fly on the GPU during
  the rendering pass and thus minimizes the amount of data transferred
  between CPU and GPU. The method directly extends to datasets that
  exceed the available graphics memory.
\item Three GPU-based rendering methods that exploit this data storage and
  access scheme, namely (1) a splatting approach that optimally places the
  splats at the mass centroids of fluid elements and locally scales 
  the kernel sizes based on the correct mass densities at these locations,
  (2) a mass conserving resampling approach, that does not suffer from problems
  of  slice-based resampling approaches which might miss parts or complete cells that
  fall between slices and (3) an efficient cell-projection approach that does not
  require any view-dependent decomposition of the tessellation.
\item A  comparison of the image quality of these new approaches to the standard
  rendering methods for N-body dark matter simulations, namely constant and adaptive
  kernel smoothing, as well as a Voronoi tessellation of the
  particle distribution. 
\item A demonstration of the effectiveness of the new approaches to 
  visualize important features of the so-called {\sl cosmic web}, in particular
  voids, filaments and dark matter halos.
\end{itemize}
}

The remainder of this paper is organized as follows. In
Section~\ref{Sec-RelatedWork} we will summarize related work in the
field of visualization of N-body simulations and direct volume
rendering of data on unstructured
meshes. Section~\ref{Sec-PhysicalMotivation} 
will review the physical motivation for the tessellation approach,
whereas Section~\ref{Sec-Rendering} focusses on the rendering
algorithm and an efficient implementation on current graphics hardware. 
Section~\ref{Sec-Results} compares the new approaches
with standard rendering methods for N-body simulations and we end with concluding
remarks and directions for future work in Section~\ref{Sec-Conclusions}.

\section{Related Work}
\label{Sec-RelatedWork} 

In the last years, numerous publications have studied the
visualization and analysis of point-based datasets from N-body and SPH
simulations and the approaches can be divided into two major categories. 

The first one comprises approaches that operate directly
on the points, e.g. by projecting kernel profiles centered at the point locations
into the frame-buffer. 
A GPU-assisted hierarchical splatting of point-based datasets via a PCA
clustering procedure has been presented and applied to various N-body
and SPH datasets by Hopft~et~al.~\cite{Hopf:2003:HSS:1081432.1081501}. Via compression and out-of-core techniques, this
work has been extended to time-dependent N-body datasets~\cite{Hopf:2004:HSS:1018014.1018059}.  An interactive rendering
approach for very large N-body datasets has been presented by
Fraedrich~et~al.~\cite{Fraedrich:2009:EMR:1638611.1639231}. 
The authors employ a continuous level-of-detail particle representation and a hierarchical
quantization scheme to compress the particle coordinates and data
attributes. A high performance parallelized algorithm for large-scale
astrophysical data sets from particle-based simulations for multicore
CPUs and CUDA-enabled GPUs has been presented~\cite{journals/procedia/JinKRGDR10}.  Popov~et~al.~\cite{Popov-UCSC-SOE-11-17}
employ the {\sl Cloud-in-Cell} method of {\sl PM} (Particle-Mesh)
schemes to resample point data onto a regular grid to
analyse so-called {\sl multistream} events, which characterize 
large-scale features like voids, halo and filaments.
Haroz~et~al.~\cite{Haroz2008}  apply multidimensional visualization
techniques to explore uncertainties of time-dependent cosmological
particle datasets. They further present a hardware-accelerated
approach for smooth temporal interpolation of the particle data in real-time.
Various other point based rendering approaches 
have been presented~\cite{szalay-2008,price-2007,Zhou06interactivepoint-based}.

  The splatting approach  proposed in this paper differs from these
  in the way the positions and sizes of the splats are computed.
  The locations are inherently coupled to properties of the
  underlying physical system, i.e. the volume conservation of 
  phase-space elements and their evolution, respectively deformation 
  over time. The locations of the splats are defined by the centers of
  mass of each of these volume elements, whereas the local splat sizes are
  directly derived from the physically correct mass density of these elements
  and not from poor isotropic density estimates based on a quite arbitrary 
  number of nearest neighbors of the tracer particles.
  This approach can be regarded as some kind of adaptive supersampling that determines 
  the location of the samples by exploiting the underlying physics of
  the data, whereas a simple, regular supersampling that does not exploit these inherent features 
  of the data would not reach the same image quality.

Work in the second category employs some kind of proxy grid, 
for example by resampling the point-based data to regular grid.
A GPU-assisted resampling approach for unstructured point data is discussed 
by Fraedrich~et~al.~\cite{Fraedrich:2010:EHV:1907651.1907951}. It 
adaptively discretizes the view-volume onto a 3D texture,  based on
the distance to the current camera position.
A GPU-assisted mapping of input particles into a volumetric
density field using an adaptive density estimation
technique that iteratively adapts the smoothing length for local grid
cells has been presented by Cha~et~al.~\cite{journals/cgf/ChaSI09}.
Another GPU-assisted resampling approach of point-data is discussed
in Zhu~et~al.~\cite{Zhu:2005:ASF:1073204.1073298}.
A method to obtain velocity field statistics from N-body
simulations by generating Voronoi and Delaunay tessellations 
has been presented by Bernardeau~et~al.~\cite{1996MNRAS.279..693B}

Our resampling approach differs from the these in the sense that it
does not operate directly on the points primitives,
but uses a tetrahedral mesh that is derived from them. The mesh
is neither a Voronoi nor a Delaunay tessellation of the computational domain,
but is rather based on the regular layout of the points that 
 N-body simulations use as initial conditions.

There has also been extensive work on the visualization of data on
tetrahedral grids. Cell-projection methods usually employ the
{\sl Projected Tetrahedra} (PT) algorithm, that decomposes each
tetrahedron into a set of triangles and assigns scalar values for the
entry and exit points of the viewing rays to each vertex~\cite{Shirley:1990:PAD:99307.99322}.
{A GPU-assisted method for decomposing the tetrahedra into
  triangles using the PT algorithm was presented by Wylie~et~al.~\cite{Wylie:2002:TPU:584110.584112}. 
}
 An artifact-free PT rendering
approach using a logarithmically scaled pre-integration table was
proposed by Kraus~et~al.~\cite{Kraus:2004:PTW:1032664.1034427}. Maximo~et~al.
developed a hardware-assisted PT approach using CUDA for visibility
sorting ~\cite{CGF29-3:903-912:2010}.  GPU-assisted raycasting methods for tetrahedral grids have,
for example, been discussed by Weiler~et~al.~\cite{Weiler:2003:HRC:1081432.1081488} and
Espinha~et~al.~\cite{Espinha:2005:HHR:1114697.1115365}.

 {We could employ these cell-projection approaches to perform the
rendering of the densities defined on our tetrahedral grid
structure. However, due to the specific problem we are focussing on,
i.~e. high-quality density projections of N-body dark matter
simulation data, we can provide a more efficient and much easier
GPU-implementation that exploits the order independency and the
implicit connectivity information given in this
case and does not require the generation of any view-dependent decompositions 
of the tetrahedra faces or any intersection computations.
}

{
An alternative method to render tetrahedral grids is to resample the
data to grid structures that are more directly supported by current
graphic hardware architectures. 
Westermann~et~al.~\cite{westermann:2001:unstructured} presented a
multi-pass algorithm that resamples tetrahedral meshes onto a
cartesian grid by efficiently determining the intersections between
planes through the centers of slabs of cells of the target grid using
the {\sl ST (Shirley-Tuchman)-classification} and {\sl OpenGL's} {\sl alpha
test} to reject fragments outside the intersection regions. 
Weiler~et~al.~\cite{Weiler:2001:HRI:601671.601702} proposed a
slice-based resampling technique to a multi-resolution grid. It
discards fragments outside the intersection regions between the slice
and the tetrahedra based on the barycentric coordinates of each
fragment, which are obtained from a texture-lookup.

The slice-based approach is problematic in our case, as it might
miss small or degenerated tetrahedra that fall between two slices. We 
need to distribute conserved quantities like the total mass of the
tetrahedron into the cells of the target grids. The resampling algorithm we
propose, estimates the volume of the intersection between the
tetrahedral elements and the cubical cells and distributes the mass
based on this information. It is easy to implement and does not require the
generation of view-dependent decompositions of the tetrahedra faces 
or additional texture-lookup tables to discard fragments outside the intersection.
}

\section{Motivation}

\label{Sec-PhysicalMotivation}
In this section we discuss the theoretical background of the
rendering methods proposed in this paper. For a more detailed 
discussion of the physics the reader might refer to~\cite{2011AbelHahnKaehler,Shandarin:2011jv}.

N-body simulations modeling the evolution of 
dark matter distributions usually discretize the computational
domain by a constant number of point-like mass sources, so-called {\sl
tracer particles}. To reduce the computational complexity, each tracer particle
represents large ensembles of physical dark matter particles, typically 
between $10^6$ and $10^9$ solar masses. 
Initial conditions are generated by distributing the tracer particles
at the nodes of a cubical grid and imposing small perturbations on
their positions and velocities according to the statistics of density fluctuations in the early
Universe, as imprinted in the {\sl CMB } ({\sl Cosmic Microwave Background})
radiation. The position of each particle $i$ is updated by
computing the aggregate gravitational forces of all other particles ${j, j\neq
  i}$ at the location of $i$, and changing $i$'s position
according to the acceleration resulting from this net force. 
In this process, the mass of the physical dark matter
particles represented by the tracers is usually treated as if it was centered
around the tracer's position.

It is important to emphasize that the tracers do not have a direct physical
equivalent, but are basically approximations introduced to keep the computational complexity 
of the simulations manageable. Even with these simplifications large-scale N-body dark matter 
simulations nowadays follow the motion of up to hundred billion tracer particles, see 
e.~g.~\cite{2009MNRAS.398.1150B,0004-637X-740-2-102}. Nevertheless, it is physically
more accurate to regard the tracer particle's mass as being spread out
over the computational domain, instead of being concentrated at a set of discrete sampling locations.

The correct time-dependent evolution  of an ensemble of dark matter
particles is given by the {\sl collisionless Boltzmann equation}, also
called the {\sl Vlasov-Poisson equation}~\cite{Peebles1993}
\begin{equation}
 \frac{ \partial f } {\partial t}  = - {\bf v} \, \nabla_{{\bf x}} f -
 \nabla_{\bf x} \, \phi
 \cdotp \nabla_{\bf v} f,   
\end{equation}
where $\phi$ is the gravitational potential of the system.
The {\sl distribution function} $f({\bf x},\bf{v},t)$ describes the phase-space
density of the ensemble, and is defined such that
$dN = f({\bf x},{\bf v},t) \, d{\bf x} \, d{\bf v}$
is the number of particles that at time $t$ have positions between
${\bf x}$ and ${\bf x+dx}$ and velocities between ${\bf v}$ and ${\bf
  v+dv}$. Given $f$, the number of dark matter
particles per unit volume $n({\bf x},t)$ at ${\bf x}$ is 
$n({\bf x},t)=\int f({\bf x,v},t) \, {\bf dv}$,
and analogously the mass density $\rho({\bf x},t)$ is
\begin{equation}
\label{Eq:Density}
\rho({\bf x},t)=  \int m f({\bf x,v},t) \, {\bf dv},
\end{equation}
where $m$ is the particle mass.

\subsection{Tessellation of the Computational Domain}
\label{Sec:Tessellation}
\begin{figure}[htb]
  \centering
  \mbox{} \hfill
 \includegraphics[width=0.975\linewidth]{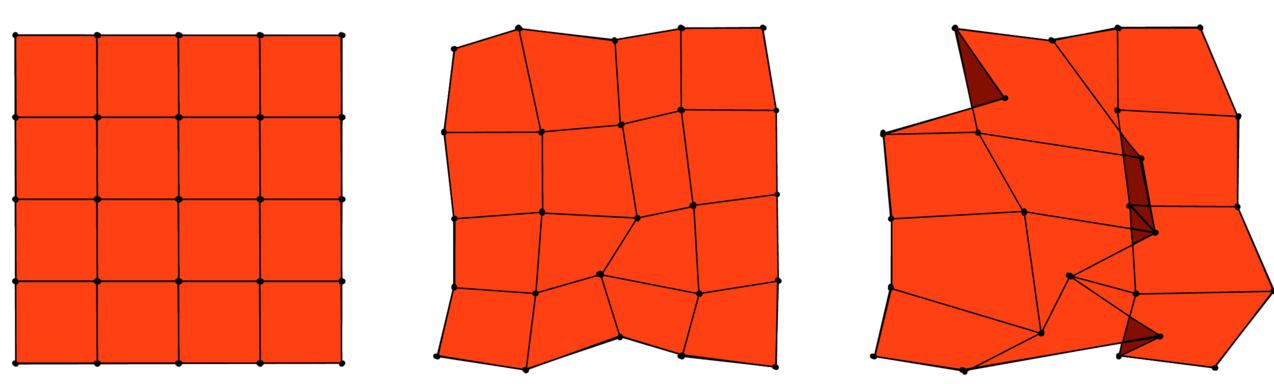}
 \hfill \mbox{}
  \caption{\label{Fig-2DMesh}%
           A 2D illustration of a regular grid structure defined by
           the tracer particles of a N-body simulation run. Initially the particles are
           distributed regularly over the computational domain
           (left). Over time the particles are advected due to
           gravitational forces and the corresponding cells become
           deformed (middle). At later times, some cells will start to 
           (partially) overlap with each other (dark shaded region on the
           right), indicating the formation of structures of the {\sl
             Cosmic Web}. }
\end{figure}

{
To illustrate how this motivates a new method to estimate the physical
quantities associated with N-body dark matter simulations, consider
the 2-dimensional phase-space diagram in
Figure~\ref{Fig-Phase-Space},  that shows the location of the fluid
elements on the horizontal axis, versus their velocities on the
vertical axis. 

\begin{figure}[htb]
  \centering
  \includegraphics[width=0.7\linewidth]{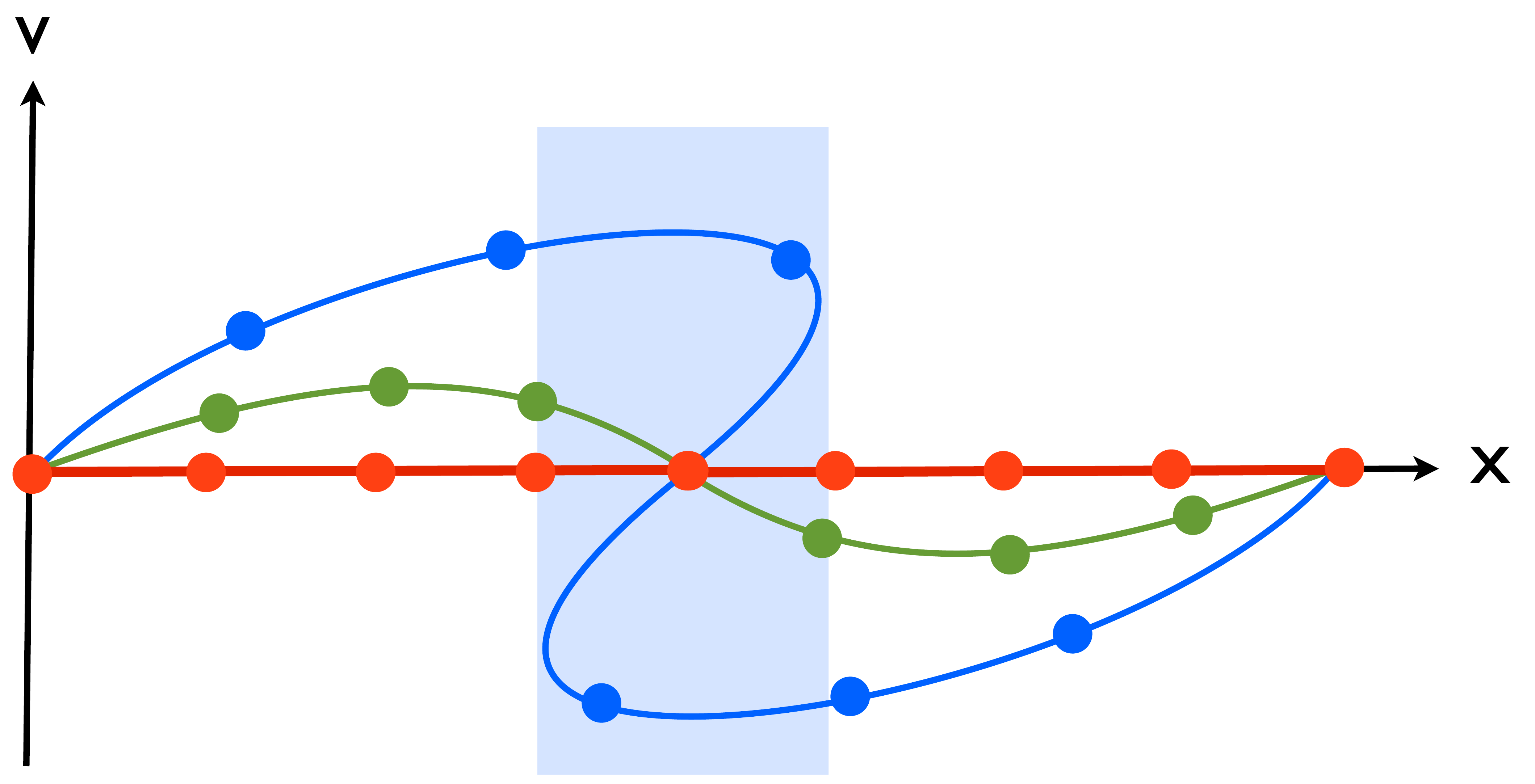}
  \caption{\label{Fig-Phase-Space}
         {This 2D phase-space diagram shows the positions and
           velocities of the dark matter fluid for three different
           time-steps, in the order of red, green and blue. At the
           latest time depicted, several
           fluid elements occupy the same location in space 
           (transparently shaded region). The dots on the lines indicate
           the location of the tracer particles used by N-body
           simulations to trace the motion of the dark matter fluid
           over time.
          }
}
\end{figure}

At early times, the dark matter fluid is almost
uniformly distributed and at rest, as depicted by the red line in 
 Figure~\ref{Fig-Phase-Space}. Over time, gravity accelerates the dark
matter fluid elements and they gain velocity, denoted by the green line. At later times different
  streams of dark matter co-exist in the same spatial regions,  in
  this example there are three regions per spatial location for elements on the blue line
  inside the transparent box.
These so-called {\sl multi-stream} regions provide
important information about the formation of structures in the dark
matter distribution on large spatial scales. The number of streams can be used
to identify regions of very low matter density, so-called voids, as well as sheets,
filaments and halos, which together form the so-called {\sl cosmic web}.
Voids correspond to regions 
with only one stream of dark matter particles, sheet-like
structures can be defined by the existence of three streams, and higher
values indicate the formation of filaments and dark matter halos, the
locations where galaxies form.

The dots on the lines correspond to the tracer particles used by the
  simulation to sample the motion of the collisionless dark matter fluid over time.}
At the initial time step,  the tracer particles are distributed uniformly in the computational
domain and their positions define the vertices of a cubical
grid in the 3D case, or squares as depicted for a 2D example in the
left image of Figure~\ref{Fig-2DMesh}. 
Since each cell initially has
the same volume and the mass
distribution is nearly homogeneous over the computational domain,
it is physically reasonable to assign a constant distribution function
$f({\bf x,v},t)$ and thus constant mass $m$
to each cell $C_i$. Over time the N-body code will
update the tracer particles according to the gravitational forces
acting in the computational domain and this will cause the initial
cubical grid cells to be deformed, as depicted in the middle image of Figure~\ref{Fig-2DMesh}.  
At later times, when gravity induced even larger inhomogeneities in the matter
distribution, the motion of the tracers leads to large numbers of cells overlapping in
the spatial domain, as shown in the right image of Figure~\ref{Fig-2DMesh}.
The crucial observation for estimating physical quantities is based on the 
conservation of mass, that states that the mass of each co-moving 
volume element is constant over time. Thus, from the knowledge of the
constant initial mass distribution and the time-dependent volume
of each cell, derived quantities like mass
densities can be computed for all times.

In principle the cubical tessellation could be employed to obtain this
information, but the non-convex cells that emerge during 
the deformation of the grid, as shown in the right image of
Figure~\ref{Fig-2DMesh}, would complicate the computation of the
time-dependent volumes. A preferable domain tessellation 
is obtained using tetrahedral elements. The advantage of
this cell type is that independently of the relative motions of the
vertices, these cells will remain convex, though the cells might
temporarily become degenerate, when all vertices (almost) lie in the
same plane. 
Tetrahedra with small volumes indicate regions of high 
mass density, since the mass per tetrahedron is constant by
construction. These high-density tetrahedra indicate caustics in the 
dark matter fluid. 

\begin{figure}[h]
  \centering
  \includegraphics[width=.4\linewidth]{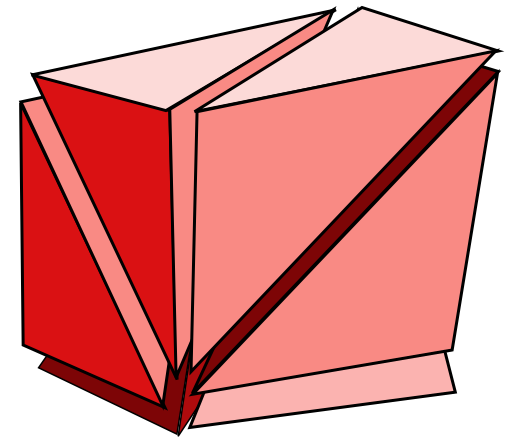}
  \caption{\label{Fig-TetraSplit}
           The decomposition of a cubical cell into six non-overlapping tetrahedra of
           constant volume used in this paper. This configuration
           introduces no new vertices besides the tracer particles of
           the simulation and ensures consistent edges and
           faces for abutting cells.
}
\end{figure}

The tessellation of the initial cubical cells should consist of
tetrahedral elements that introduce no new vertices, ensure consistent
faces and edges between abutting cells, and initially have 
identical volumes.
The smallest number of elements that fulfill these constraints is six, and we chose
the configuration shown in Figure~\ref{Fig-TetraSplit}.  
 This choice ensures that no holes or cracks will form in the
interior region of the mesh over time, even as the grid gets vastly deformed, 
because the vertices and edges on shared faces match up.

The connectivity that defines the tetrahedra is kept the same for
all time steps. Only the spatial positions of the vertices are updated 
according to the actual positions of the tracer particles, as 
computed by the N-body simulation. This implies that for a new time step,
only the positional information of the vertices must be updated, 
while the connectivity information can be reused. The identification of
corresponding particles for different time-steps is done with help
of the unique IDs that simulation codes assign to the tracers.
These can be mapped to the coordinates of the particles on
the initial grid, see Equation~(\ref{Eqn:ID}).

\subsection{Density Projections}

According to the discussion in Section~\ref{Sec:Tessellation}, the local mass density
of a co-moving tetrahedron $i$ at time $t$ is given by
 \begin{eqnarray}
\label{Eqn-rho_loc}
 \rho_{loc,i}(t) = \frac{m}{V_i(t)},
\end{eqnarray}
i.e.~the constant mass $m$ divided by its time-dependent volume
$V_i(t)$. Over time, the motion of the tracer particles
results in large amounts of overlapping volume elements, and according
to Equation~(\ref{Eq:Density}) the total mass density $\rho_{tot}(\mathbf{x},t)$  at a position
$\mathbf{x}$ is simply given by the sum of the densities of each
cell containing $\mathbf{x}$ at that time. To illustrate this, consider the
volume element $V_\cap(\mathbf{x},t)$ that is obtained by intersecting all cells
that contain $\mathbf{x}$ at time $t$. The total density
is then defined by
\begin{eqnarray}
\label{Eqn-rho_tot_a}
 \rho_{tot}(\mathbf{x} ,t) = \frac{\sum_{i} m_{i,\cap} }{V_\cap(\mathbf{x},t)}.
\end{eqnarray}
where the sum runs over all cells that contain $\mathbf{x}$, and $m_{i,\cap}$ is
the amount of mass per cell $i$ contained in subvolume $V_\cap$. Using constant spatial
interpolation, we get $m_{i,\cap}  = m \frac{V_{\cap}}{V_{i}(t)}$,
where $m$ is the constant mass per cell and $V_i$ is the cell's volume. 
Combining this with Equation~(\ref{Eqn-rho_tot_a}) we get 
 \begin{eqnarray}
   \label{Eqn-rho_tot_b}
 \rho_{tot}(\mathbf{x} ,t) = \sum_{i} \rho_{loc,i}(t),
\end{eqnarray}
where the sum runs over all cells $i$ that contain $\mathbf{x}$
at time $t$. The number of dark matter streams at a certain spatial location,
which as discussed above can be used to distinguish different regions of the
cosmic web, corresponds to the number of overlapping tetrahedra.

Relevant for many scenarios in astrophysics and cosmology are projections of 
certain physical quantities $q$ along the line of sight 
\[
q_{proj} = \int_{\chi} q \big(\mathbf{x(\chi}) \big) d\chi,
\] 
where $\mathbf{x(\chi)}$ denotes the parametrization of that line for
a certain pixel on the screen. For the tetrahedral mesh discussion above,
it takes the form
\begin{equation}
  \label{Eqn:TetProjection}
  q_{proj} = \sum_i \bar{q}_i dl_i,
\end{equation}
where the index $i$ runs over all tetrahedra $T$ that are intersected by
the line of sight, $\bar{q}_i$ is the constant quantity associated with
$T_i$ and $dl_I$ denotes the length of the intersections between the
line of sight and $T_i$. Particularly important are density projections
\begin{eqnarray}
\label{Eqn-column-density}
  \rho_{proj} = \int_{\chi} g \big(\rho(\mathbf{x(\chi}) \big) d\chi, 
\end{eqnarray}
where $g$ is some function of the density $\rho$.
$n=1$ for example is relevant for experiments aiming at detecting dark matter directly in
underground detectors. 

The discussion in this section can be summarized as follows:
Given a time-dependent 3D N-body dark matter simulation, a tetrahedral
mesh is constructed, with a connectivity implicitly defined by the layout of the
tracers on a regular grid at the initial time-step, which can be
reconstructed at any time step from the tracer's unique IDs.
The same amount of mass is assigned to each tetrahedral element 
and derived quantities, like time-dependent mass densities, are
computed based on the volumes of all tetrahedral elements that 
overlap a certain location, see Equation~(\ref{Eqn-rho_tot_b}). The
mass is associated with the cells and not the vertices of the tessellation.
The nodes of the mesh are updated over time, according to the tracer's actual
position, changing the volumes and thus the spatial mass densities.
The tessellation has consistent vertices, edges and faces for
abutting cells, and in particular does not contain any dangling
nodes, but at later times the tetrahedral elements will typically
start to overlap.

\section{Rendering}
\label{Sec-Rendering}

\begin{figure*}[htb]
  \centering
  \includegraphics[width=0.995\textwidth]{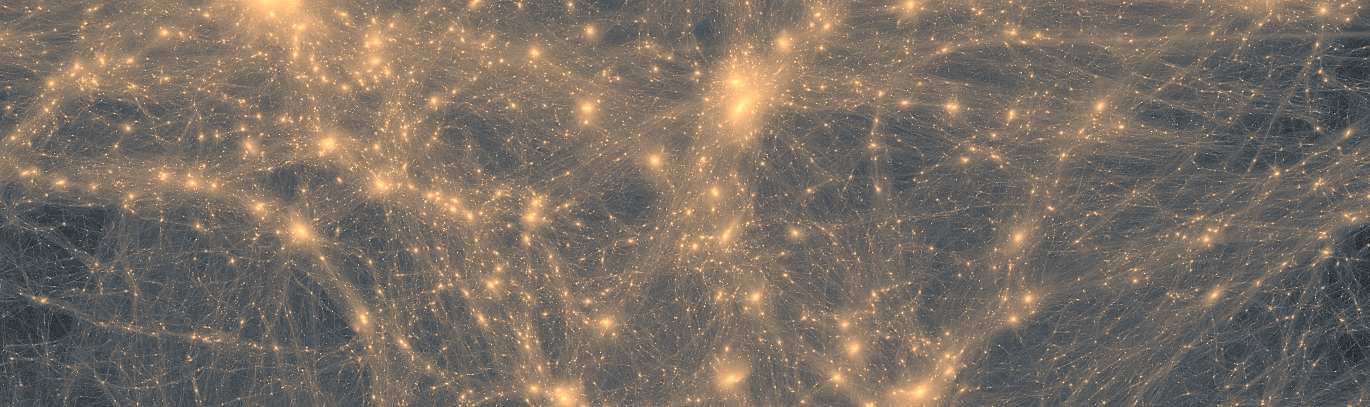}\\
  \vspace{0.2cm}
\includegraphics[width=1.\textwidth]{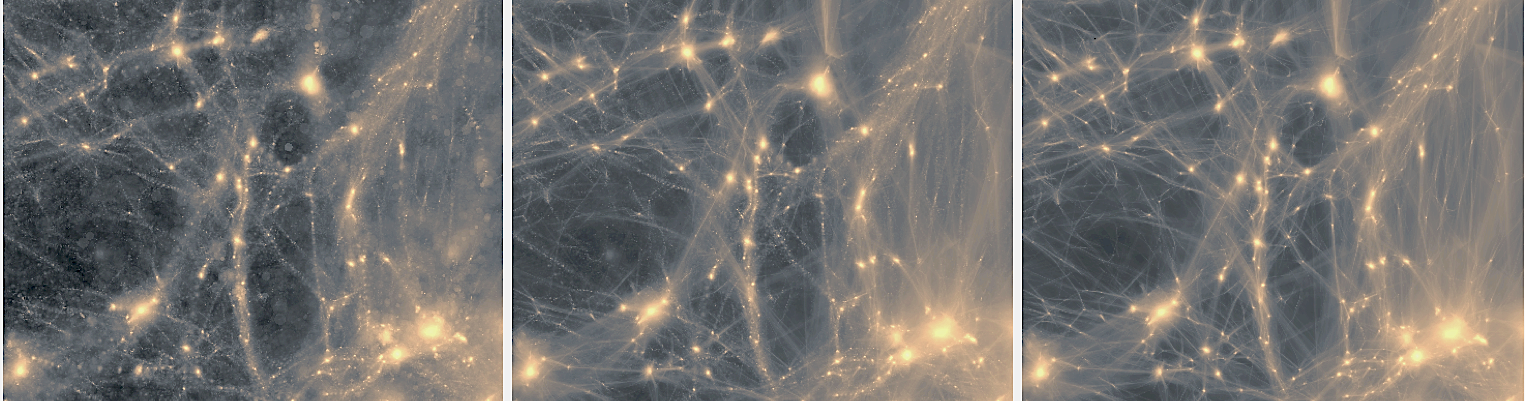}
\caption{A density distribution from a dark matter simulation
  with about 134 million particles, resulting in about 804 million
  tetrahedra, respectively 3.2 billion triangles. The close-up on the
  left was rendered via the centroid approach ($0.1$~fps), whereas the 
  image on the right was generated with the 
  cell-projection approach ($0.03$~fps). For the image in the middle a hybrid approach was used,
  rendering the tetrahedral elements inside the sphere around the
  camera using the cell-projection methods and elements outside a
  sphere with the centroid
  methods ($0.08$~fps).}
\label{Fig-LargeScale}
\end{figure*}

In the following we discuss three GPU-based rendering approaches
for density projections generated from this type of input mesh.
We implemented them in {\sl OpenGL} and the {\sl OpenGL Shading
  Language}, and we will use {\sl OpenGL} nomenclature in
the following.

\subsection{Data Storage and Access on the GPU}
\label{Sec::DataStorage}
The relation between a tracer's ID and its vertex $(i,j,k)$ on
the initial regular grid is given by
\begin{equation}
id = i + d_0(j+d_1 k) \label{Eqn:ID}
\end{equation}
where $d_0,d_1,d_2$ are the number of vertices of the grid along the $x$, $y$
and $z$ direction. The implicit connectivity of the initial grid allows for a very memory efficient
representation of the mesh on the GPU without the need to store and transfer any explicit
connectivity information or additional attributes about the tetrahedral
cells, except for the locations of the tracer particles. All connectivity and derived
information, like the volumes and mass densities of the tetrahedral
cells can be generated on-the-fly on the GPU. 

The tracer particle positions are stored in a three-dimensional  floating-point
RGB texture with $d_0 d_1 d_2$ texels, so that several particles can be
accessed in the vertex shader instance. The texel coordinates $(i,j,k)$ are
derived from the particle's ID, according
to Equation~(\ref{Eqn:ID}). The texture is uploaded onto the GPU and sampled
in a vertex shader, which is invoked $(d_0-1)*(d_1-1)*(d_2-1)$ times
via instance rendering ($glDrawElementsInstanced(...)$).
The current invocation ID is obtained from the
instance counter ({\sl gl\_InstanceID}) in the vertex shader mapped to
texel coordinate $(i,j,k)$, see Equation~(\ref{Eqn:ID}).
The coordinates ${\bf v_i}, i=0...7$ of the eight tracers stored at $(i \pm 1, j \pm 1, k \pm 1)$, defining
a cubical cell in the initial regular grid, are read from the 3D
position texture and handed over to a geometry shader as varying attributes. 
In the geometry shader, the six (possibly deformed) tetrahedra depicted in
Figure~\ref{Fig-TetraSplit}, are constructed from the tracer's
positions via the connectivity table
$\{1,0,2,4\}$, $\{3,1,2,4\}$, $\{3,5,1,4\}$, $\{3,6,5,4\}$, $\{3,2,6,4\}$, $\{3,7,5,6\}$,
and the volume of each tetrahedron is computed by
\begin{equation}
  \label{Eqn:TetVolume}
  V = \frac{ ({\bf v_1-v_0} ) \cdot (({\bf v_2-v_0})\times({\bf v_3-v_0})) }{6}.
\end{equation}
In order to better leverage the massive parallelism of current GPU
architectures, we do not generate all six tetrahedra in the same
geometry shader instance, but rather
trigger six geometry shader invocations for each vertex shader
instance via the ``invocations'' layout qualifier of the {\sl OpenGL
Shading Language} ``{\sl layout ( points, invocations = 6 ) in;}''.
The built-in variable $gl\_InvocationID$ is used to determine
which of the six tetrahedra to be generated in a certain geometry shader instance.

{
For datasets that exceed the available graphics memory, we
decompose the 3D texture that stores the positional information 
into separate blocks, each of them small enough to fit
entirely into graphics memory. The blocks share a layer of texels at
their interfaces, and are transferred and processed individually on the
GPU, simply accumulating the partial rendering results for the
tetrahedral elements encoded in each brick.
}

In the following subsections, we will discuss three different
GPU-assisted rendering approaches that are based on this data storage 
and access strategy.  We will focus on the rendering of density
projections, see Equation~(\ref{Eqn-column-density}), which  are
order-independent,  so no sorting of the rendering primitives is required.

\subsection{Centroids}
\label{SubSec:Centroids}
In this approach, each tetrahedron $T$ is rendered using a 2D billboard
with a cubic-spline kernel, oriented perpendicular to the
current viewing direction and located at $T$'s
centroid ${\bf c}_T = \frac{1}{4} \sum_{i=0}^{3} {\bf v}_{i}$,
 where the $v_{i}$ denote $T$'s vertices. Since we are assuming constant mass 
(density) per cell, the centroid is identical to $T$'s center of mass.
The kernels are scaled proportional to $\sqrt[3]{ \frac{m}{\rho_T}}$,
where  $\rho_T$ denotes the mass density of $T$, that is computed in
the geometry shader according to Equation~(\ref{Eqn:TetVolume}), along with
the vertices and texture coordinates for the billboards. The
contribution of each generated fragment is accumulated
in a floating-point 2D texture that is bound as a render target,
using an additive blending equation.

The centroids and the sizes of the quadrilaterals could be computed in a
preprocessing step and cached on the GPU using {\sl vertex buffer objects (VBOs)}. 
However, since we have about $6$ times more centroids than tracer
particles, storing the centroids along with the density-dependent scaling factors
would result in a considerable increase of bandwidth and graphics memory
consumption. It is thus preferable to generate this information
on-the-fly on the GPU.

\subsection{Resampling}
\label{SubSec:Resampling}
\begin{figure}[]
  \centering
  \mbox{} \hfill
 \includegraphics[width=0.5\linewidth]{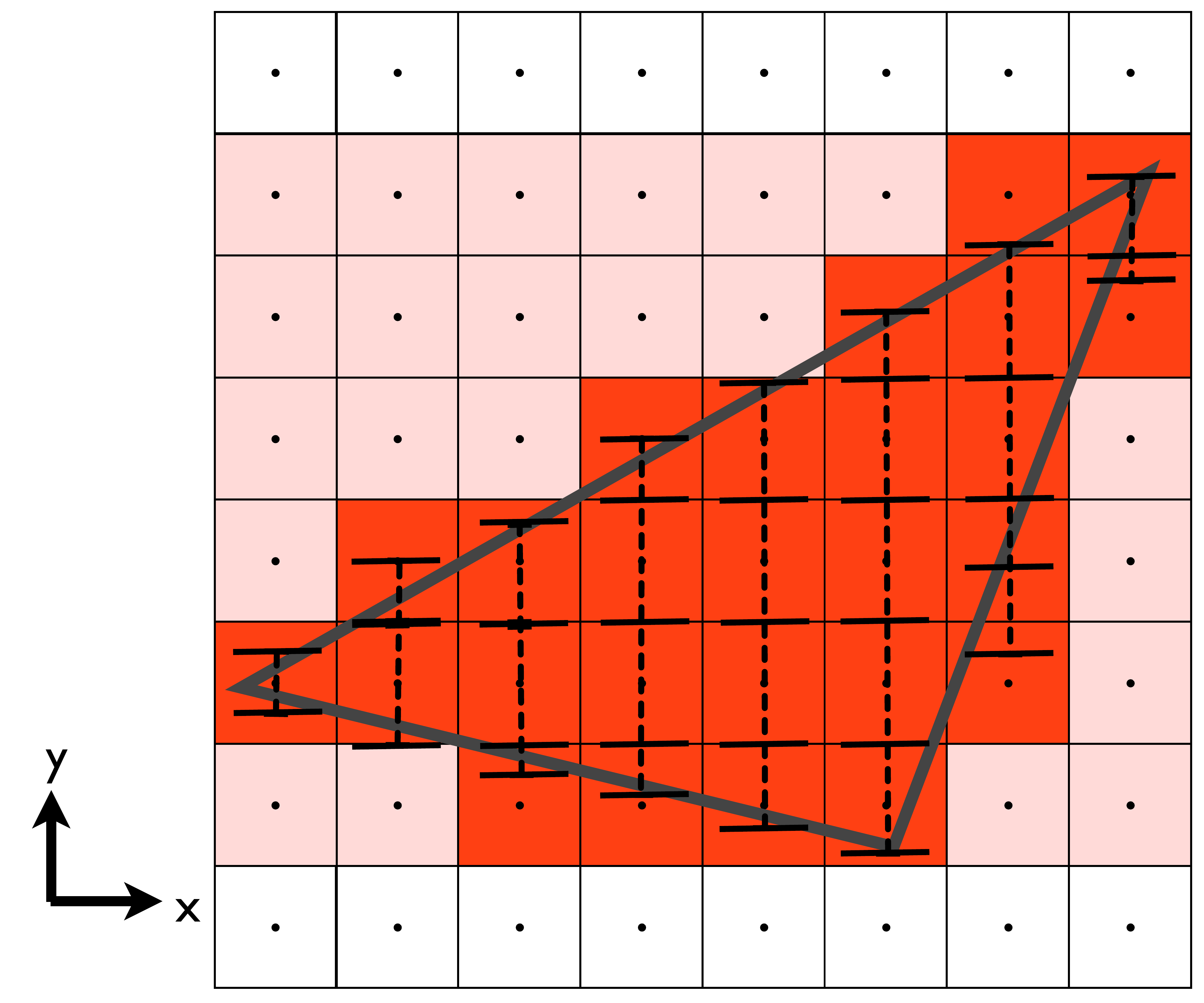}
 \hfill \mbox{}
 \caption{\label{Fig-Resampling}
   2D-illustration of the resampling approach: The cubical grid is processed
   in slabs of cells along the x-direction. The shaded cells are 
   affected by the projection of the tetrahedron (triangle) onto the
   slabs along the y-direction and thus two fragment per cell are generated. 
   For light-red cells the contribution of the fragments for front and back-faces
   cancel out, for cells that intersect the tetrahedron, the
   resulting line-segments $dl$ of Equation.~(\ref{Eqn:ResResult}) are shown.
}
\end{figure}

The second approach we propose is to resample the tetrahedral cells to a cubical
grid structure $G$, which allows to locally evaluate Equation~(\ref{Eqn-rho_tot_a}) and to
apply standard rendering methods for regular grids, e.g. to
display level sets of the data. {The mass of each tetrahedral
element $T_i$  needs to be distributed to the cubical cells $C_j$ of $G$. 
A correct solution involves the computation of the 
volume of the intersections between $T_i$  and the cubical cells $C_j$ and
an assignment of the mass contribution based on its volume, see
Equation~(\ref{Eqn-rho_loc}). For the hundreds of millions of
tetrahedra in typical N-body simulations, this procedure
is too expensive and we utilize rasterization hardware to estimate
the amount of mass for each cell by  $A \; \rho_i \; dl_i$. $A$ is the 
area of the faces of the cubical cell, $\rho_i$ is the mass
density of $T_i$  and $dl_i$ is the height of the
intersection between $T_i$ and $C_j$ at $C_j$'s center, measured along the grid's $z$-axis,
see~Figure~\ref{Fig-Resampling}.
}

We exploit {\sl OpenGL}'s {\sl render-to-texture} 
functionality and bind a floating-point 3D texture with $n^3$ texels as a
render target, using a viewport size of $n^2$ pixels. 
The tetrahedral cells and associated information, such as 
their volumes and densities, are generated in the vertex and
geometry shader as described in Section~\ref{Sec::DataStorage}.
In the geometry shader we determine the interval of $G$'s cell slabs 
perpendicular to the $z$-direction, that is partially overlapped by the
tetrahedron, see Figure~\ref{Fig-Resampling}.
Using parallel projection along the $z$-axis, the faces of the
tetrahedron are rendered to corresponding texels in the 
render target, which is specified by the {\sl gl\_Layer} command. 

We assign the derived densities,
as well as the current position of the slab of cells as varying variables to
the vertices of the triangles, to access this information in the fragment shader stage.
The tetrahedral cells can extend over many cell slices in the
$z$-direction, so that the number of generated vertices per geometry shader
instance could in principle exceed the limit of the graphics hardware. 
We therefore perform a multipass approach and process a fixed number
of cell slabs in $G$ per pass, ensuring that the maximal possible number of vertices per
geometry shader invocation stays within the valid limits.

In the fragment shader, the fragment's $z$-coordinate $f_z$ is compared
to the minimal and maximal z-coordinates $s_{min}$, resp.~$s_{max}$ of
the current slab of cells in the target texture. If the fragment belongs
to a front-facing triangle, the value
\begin{equation}
\label{Eqn:ResFF}
  q_{ff} = -q_i min \, \bigl( s_{max}, max( f_z, s_{min}) \bigr)
\end{equation}
is written to the corresponding target cell in $G$.
Analogously,
\begin{equation}
\label{Eqn:ResBF}
  q_{bf} = +q_i \, max \bigl( s_{min}, min( f_z, s_{max}) \bigr)
\end{equation}
is written for back-facing fragments. The different signs in Equation~(\ref{Eqn:ResFF}) and
(\ref{Eqn:ResBF}) guarantee that contributions of fragments that correspond to
target cells outside the tetrahedron, shaded in light-red in Figure~\ref{Fig-Resampling},  will
cancel out. The resulting net value written to the cell $C$ in $G$ is
\begin{equation}
\label{Eqn:ResResult}
  q = q_{bf} + q_{ff} = 
  \left\{ \begin{array}{rcl}
         &    0 & \mbox{if  }  C \cap T = \emptyset,  \\
         &q_i dl & else  
  \end{array} \right. ,
\end{equation}
where $dl$ denotes the height of the intersection between $T$ and
$C$ at the $C$'s center, along the $z$-direction. 

After all tetrahedra are processed, each cell
stores a local approximation of Equation~(\ref{Eqn-column-density}).
The resulting 3D texture is kept on the GPU and can, for example, be 
rendered using a standard GPU-based ray-casting approach.

\subsection{Cell-Projection}
\label{SubSec:Cell_Projection}
{The third rendering method evaluates Equation~(\ref{Eqn:TetProjection})
with a cell-projection method, computing the contributions 
$\bar{q}_i dl_i$ for each tetrahedron $T_i$. 
Rewriting Equation~(\ref{Eqn:TetProjection}) as

\begin{eqnarray}
q_{proj}
&=& \sum_i \bar{q_i} | \mathbf{b}_{i}-\mathbf{f}_{i}| \nonumber\\
&=& \big( \sum_i \bar{q_i} | \mathbf{b}_{i}-\mathbf{c}| \big) - \big( \sum_i \bar{q_i} | \mathbf{c}-\mathbf{f}_{i}| \big) \label{Eqn:TetSum},
\end{eqnarray}

where $\mathbf{f}_{i}$ and $\mathbf{b}_{i}$ denote the entry and exit
points of the line of sight for $T_i$ and $\mathbf{c}$ is the current camera
location shows that Equation~(\ref{Eqn:TetProjection}) can be evaluated efficiently by separately
adding the contributions of the front-facing and back-facing triangles.
}

As in the previously discussed approaches, the vertices ${\bf v}_i$ of each deformed cell are
obtained by sampling the position texture in the vertex shader and
the six tetrahedra are constructed in the geometry shader.
The faces of the tetrahedra are rendered as triangle strips.
A negative value of the volume formula Equation~(\ref{Eqn:TetVolume}) indicates 
that the tetrahedron is inverted, and the order of vertices in the
strip has to be adjusted to ensure consistent face orientations.
The mass densities are computed from the volumes and handed over to
the fragment shader as  varying variables.

In the fragment shader, the contributions $\bar{q}_i dl_i$ are computed for each tetrahedron $T_i$. 
Therefore, the blending equation is set to the additive blending equation $C_{src}+C_{dest}$,
and the fragment shader stores the contributions 
for fragments of front-facing triangles
$\sum_i \bar{q_i} | \mathbf{f}_{i}-\mathbf{c}|$ 
in the {\sl red} channel of the frame-buffer,  and the contributions for
back-facing fragments
$\sum_i \bar{q_i} | \mathbf{b}_{i}-\mathbf{c}|$
in the {\sl green} channel. 
After the triangles for all $T_i$ are processed, a separate fragment
shader computes the final sum in Equation~(\ref{Eqn:TetSum})  by 
subtracting the partial sums that are stored in the red and green
channels and the result for each pixel is written into the frame-buffer.

\section{Results}
\label{Sec-Results}

\begin{figure*}[htb]
  \centering
\includegraphics[width=1.\textwidth]{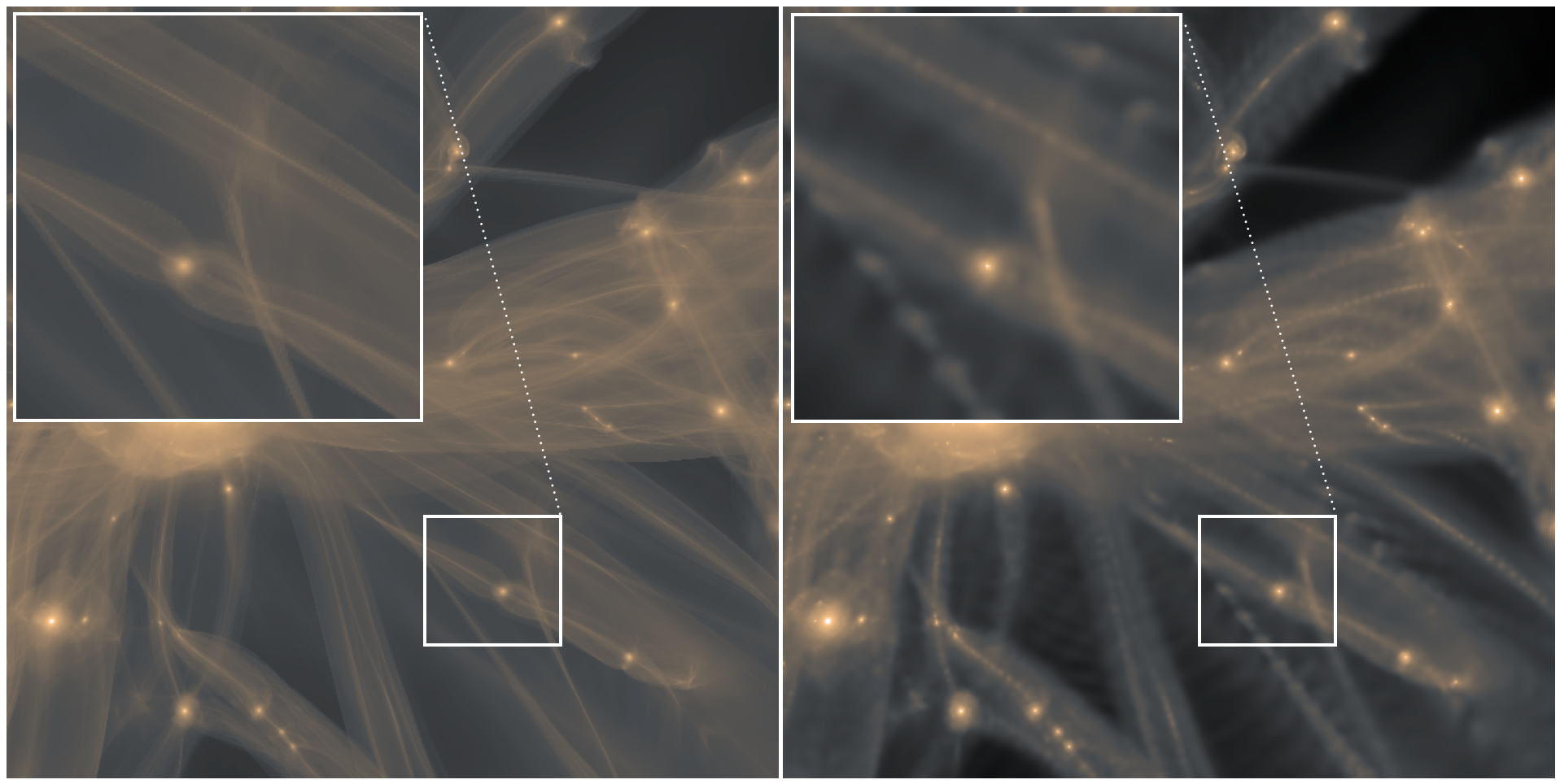}
\caption{\label{Fig-TeaserComp}%
 A direct comparison between our tetrahedral cell-projection approach
 (left) and a standard SPH adaptive kernel smoothing method. Artifacts 
 due to the poor density estimates in
 low-density regions are obvious for the SPH method, whereas the
 tetrahedral approach achieves an overall high image quality, on small
and large structures.}
\end{figure*}

The comparison was performed using  
a {\sl Nvidia Quadro 6000} graphics card with $6$
GByte of graphics memory, that was installed on a host with a 
 {\sl Intel Xeon E5520 CPU} and 24 GByte main memory.
The rendering algorithms were implemented 
in {\sl OpenGL} and the {\sl OpenGL Shading Language}. 

\begin{figure}
 \centering
\includegraphics[width=0.475\textwidth]{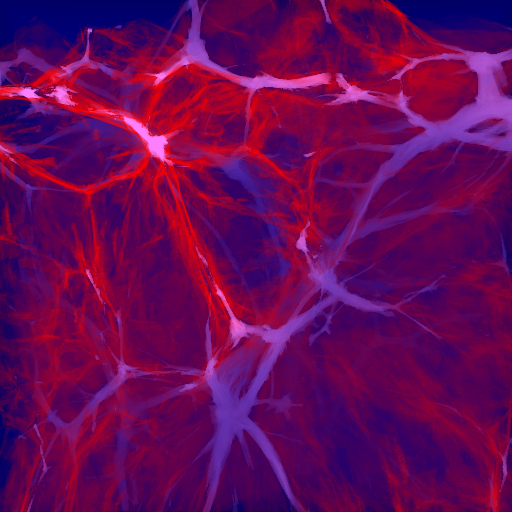}
\caption{\label{Fig-Multistreams}%
 A visualization of the so-called {\sl multi-stream} field that
 shows the number of dark matter streams at each location. Voids are
 shown in blue, sheet-like structures in red and filaments in white. }
\end{figure}

Figure~\ref{Fig-LargeScale} shows a rendering of a dark matter
simulation with 134 million tracer particles using the cell-projection
approach from Section~\ref{SubSec:Cell_Projection}. 
Figure~\ref{Fig-Multistreams} shows a visualization of the {\sl multi-stream} field, that
 counts the number of dark matter streams at each location in the
 computational domain. The data were resampled to a $512^3$ grid,
 setting $q_i=1$ in Equations~(\ref{Eqn:ResFF}) and (\ref{Eqn:ResBF}),
 to count the number of
 tetrahedra per cell, and rendered using a GPU-assisted ray-caster. Voids,
 shown in blue, sheet-like structures (red) and filaments (white), can
 be clearly distinguished.

\begin{figure*}[]
  \centering
\includegraphics[width=0.475\textwidth]{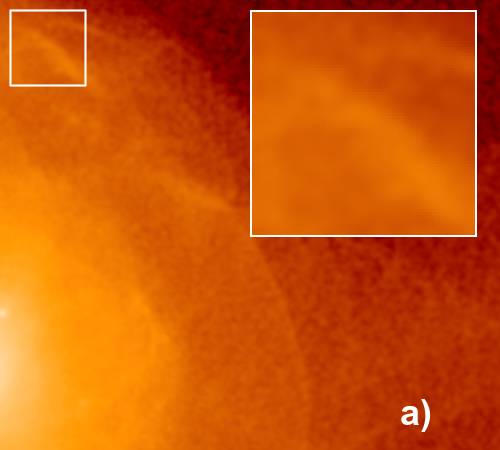}
\includegraphics[width=0.475\textwidth]{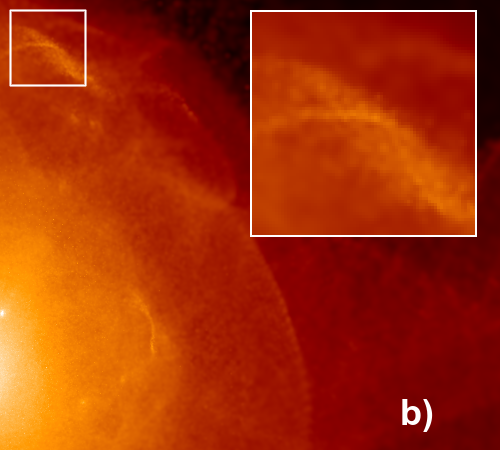}
\includegraphics[width=0.475\textwidth]{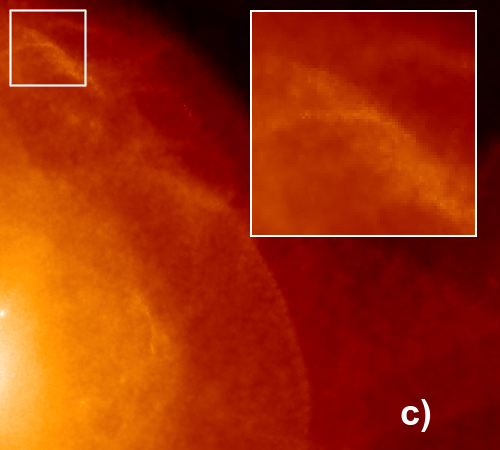}
\includegraphics[width=0.475\textwidth]{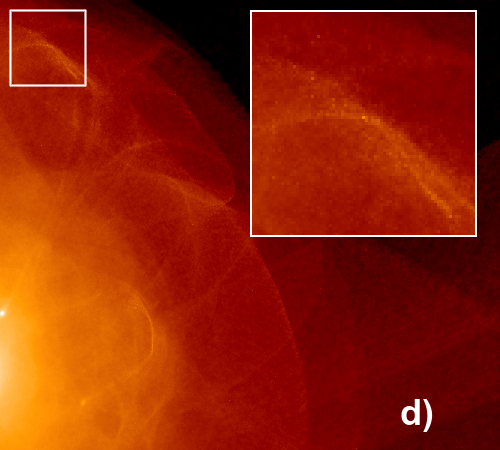}
\includegraphics[width=0.475\textwidth]{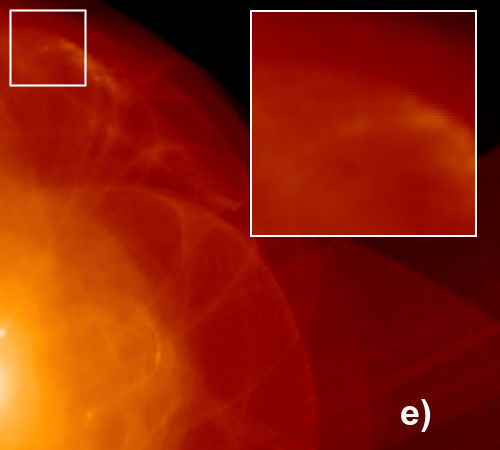}
\includegraphics[width=0.475\textwidth]{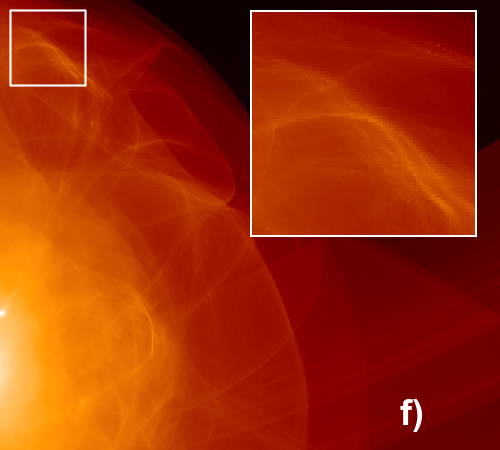}
  \caption{ Comparison between the visualizations of a dark matter halo
    simulation using three conventional techniques, namely {\sl
      constant kernel smoothing} (a), {\sl adaptive kernel
    smoothing} (b), {\sl voronoi tessellation} (c), and the three new rendering
    methods based on the tetrahedral phase-space tessellation proposed
    in this paper, i.~e. {\sl centroids} (d), {\sl resampling} (e) and
    {\sl cell-projection} (f). The subimage on the right shows a close-up of the
    rectangular region of the left. }\label{Fig:Comparison}
\end{figure*}

We further compared the image quality and performance of the three 
rendering methods proposed in Section~\ref{Sec-Rendering}
with three conventional approaches for density projections from a dark matter N-body simulation
with 17.2 million particles, see Figure~\ref{Fig:Comparison}.
The screen size was $1460$x$860$ pixels.
{In Figure~\ref{Fig-Multistreams} an emission-absorption scheme was
chosen as the lighting model. In all other rendering
examples shown in this paper, the resulting total mass density in the
framebuffer was logarithmically rescaled in a separate fragment
shader and mapped to colors via a 1D-texture lookup.
}

Close-ups of the resulting images are shown in Figure~\ref{Fig:Comparison}.
For images~a) and b), the tracer particles were rendered using cubic
spline kernels, 
accumulating the contributions in the frame-buffer. The points were cached on the GPU
using {\sl vertex-buffer arrays} and the geometry for the view-port
aligned billboards was generated in the geometry shader. For image~a),
constant kernel sizes were used for all particles. For image~b),
adaptive kernel sizes based on local density estimates
were projected. These were obtained from the smallest spheres
enclosing the $32$~nearest neighbors 
around each particle, a standard approach in SPH simulations, as for
example discussed by Springel~et~al.~\cite{Springel05thecosmological}.
The sizes of the kernels were scaled proportional to $\rho_i^{-1/3}$,
where $\rho_i$ denotes the resulting density estimation for particle
$i$. The adaptive kernel sizes were computed on the CPU using a
kd-tree search tree and cached on the GPU
along with the positions using {\sl vertex-buffer arrays}. 

For image~c), the Voronoi tessellation of the $17.1$ million dark matter tracer
particles was generated using the {\sl Voro++}~library~\cite{voro++}.
The density around each particle was
computed from the volume of its Voronoi cell. The resulting density 
projection was generated via a cell-projection approach. Therefore, the
cell faces were rendered separately in the $GL\_POLYGON$ mode. In the 
fragment shader, the fragment's distance $d$ to the camera location was computed and 
$d \, \rho$ was written to the red-channel for front-facing fragments, and
respectively to the green-channel for back-facing fragments. After all
cells were processed, the difference between the red and green
channels was written to the image buffer, yielding the line integral
of the density, as discussed in Section~\ref{Sec-Rendering}.
We were primarily interested in the image quality and did not optimize the
rendering performance for the special case of Voronoi cells.
Approaches like the one discussed by Muigg~et~al.~\cite{muigg-2011-gpg} would certainly perform 
much faster, though problems would arise
due to  the large amount of geometry and connectivity information that
has to be encoded in the 3D textures for the considered Voronoi mesh.   
Image~d) in Figure~\ref{Fig:Comparison} shows the rendering result for the same
dataset and camera position using the centroid method described in
Section~\ref{SubSec:Centroids}. To ease comparison, we have used the
same cubic spline profile for the six times more numerous tetrahedron
centroids.
Image~e) was generated by resampling the $6*17.1=102.6$ million
tetrahedral elements onto a regular grid with $512^3$ cells using the
resampling approach discussed in Section~\ref{SubSec:Resampling}. The resulting grid
was rendered using a standard GPU-raycasting approach. 
Finally, image~f) in Figure~\ref{Fig:Comparison}  was generated via
the cell-projection approach for the $102.6$ million tetrahedra, as
discussed in Section~\ref{SubSec:Cell_Projection}. 
Figure~\ref{Fig-TeaserComp} shows another direct comparison between
the tetrahedral cell-projection approach  and the SPH adaptive kernel 
smoothing method.
An overview about the preprocessing times, the memory requirements and
performance numbers are summarized in Table~\ref{Fig-Table}.

\begin{table}
\caption{Memory requirements and rendering performance 
   of the 6 different methods, namely constant kernel smoothing~(a),
  adaptive kernel smoothing~(b), Voronoi tessellation~(c), and the
  three new rendering methods based on the tetrahedral phase-space
  tessellation proposed in this paper, centroids~(d), resampling
  (e)~and cell-projection~(f).} \label{Fig-Table}
\begin{tabular}{|r||r|r|r|r|r|r|}
\hline
 & a & b & c & d & e & f\\
\hline 
memory [GBytes] & 0.20 & 0.26 & 2.80 & 0.20 & 1.7 & 0.20\\
preprocessing [s] & 0.2 & 185 & 962 & 0.2 & 52 & 0.2  \\ 
performance [fps] & 3.0 & 3.0 & 0.001 & 0.5& 2.2 & 0.1 \\
\hline
\end{tabular}
\end{table}

\section{Discussion}
\label{Sec:Discussion}
The images in Figures~\ref{Fig-TeaserComp} and~\ref{Fig:Comparison}  clearly demonstrate the
improved image quality of the new rendering methods, as depicted in images
d) to f). Especially image~f) and Figure~\ref{Fig-TeaserComp} show that
the proposed cell-projection approach  achieves very high
image quality, both in areas of homogeneous densities, for example in the filaments emerging from
the central halo, and at the same time reveal significantly more
fine-scale details in the central region of the dark
matter halo. Caustics,
formed at the locations where orbits of the dark matter fluid turn
around, become clearly visible. At the same time, the centers of
filaments and their inner and outer caustics become obvious.
Filamentary and sheet like structures connecting dense knots
(dark matter halos) are more easily recognized.
The standard methods shown in images a) to c) suffer from rendering 
artifacts due to poor isotropic density estimates, resulting in quite high image
noise. The noise for a) and
b) could be reduced by increasing the overall kernel sizes, but this
would result in an increased smoothing of the fine-scale details in the
central region of the halo. 

However, the superior image quality of the cell-projection approach
comes at the cost of a lower rendering performance compared to the
point-based rendering methods in a) and b). Here the centroid
and the resampling approaches offer a good trade-off, achieving still
much better image quality than the traditional methods, while achieving 
comparable rendering performance as the point-splatting methods shown 
in images~a) and b) of Figure~\ref{Fig:Comparison}.
Because the algorithms we propose generate all
connectivity information and derived fields
like mass density on the GPU on-the-fly, the memory resources and 
amount of data that need to be transferred between CPU and GPU are 
considerably smaller compared the standard  Voronoi tessellation.

\subsection{Scientific Insights}
The images in Figure~\ref{Fig-TeaserComp} clearly demonstrate the
superior image quality of the new rendering method. Perhaps even more
interestingly is that it gives a deep insight in the origin of the artificial
numerical fragmentation found in N-body techniques for more than 30
years, see \cite{Wang:2007a}, and references therein. As the particles 
evolve and one thinks of them as regions that in some spherical kernel 
contain their mass, filaments become artificially clumpy. In the
simulation itself, this artifact leads to a potential well which further
deepens as the particles attract each other and accrete more particles 
from their surrounding. Consequently, entire halos are found to
originate purely from errors made in the gravitational forces by
assuming that the particles are isotropically softened gravitational point masses. 
Remarkably, the approach we use here is physically accurate and 
shows directly where such incorrect clumps originate. Our full
tetrahedral projection method has the filaments shown in
Figure~\ref{Fig-TeaserComp} to be perfectly smooth 
and bounded by the caustics formed by particles orbiting in the
filament potential and shows no sign of artificial clumping. From
these images the domain scientists are given the important insight, 
that if they can construct solvers which use our density field to
compute the gravitational potential from they will very likely be able
to avoid these undesired numerical artifacts which have hindered
reliable studies of the halo mass function in warm and hot dark matter 
models as well as the understanding of how in detail the very first dark
matter halos form in the standard Cold Dark Matter model.
Furthermore, for the first time our method allows reliably to check
for the existence of caustics in the DM density distribution for simulations already run.

The domain scientists have further
benefited from being able to extract two dimensional slices through the
data using the proposed methods. Previously, only full or partial projections had been shown, but 
being able to measure the density, velocities, etc. on infinitesimally 
thin slices had been missing. This capability allows also for a
much closer visual comparison with the hydrodynamics properties 
of the gaseous matter typically evolved at the same time as the dark 
matter in the most sophisticated computational calculations.

The images generated with the new rendering methods can be 
directly used as input for predictions of the gravitational lensing
effect (cf. e.g. \cite{Hilbert:2009}). Images of the mass density directly correspond to so-called
convergence maps, but also so-called shear maps can be computed
in a straight-forward way once an image is at hand. The clear
advantage over previous approaches is the low noise level of
our images that does not come at the price of a large isotropic
filtering that washes out relevant small-scale structure. We are
currently working on using the rendered images for this purpose.

\subsection{Scalability to Large Data}
As discussed in Section~\ref{SubSec:Resampling}, the rendering methods
presented in this paper also extend to datasets that exceed the
available graphics memory. In this case,  the 3D texture used 
to store the positional information is decomposed into separate sub-bricks, 
each of them small enough to fit entirely into graphics memory. 
It would be straight-forward to apply this technique to run the
algorithms on a GPU-cluster, by distributing the separate bricks to
the individual cluster-nodes. Each brick could be processed in
parallel and the partial density projections would be added to obtain the final rendering result.
The choice between the three different methods allows
for a trade-off between performance and image
quality for example by choosing the high quality cell-projection
methods for regions close to the camera and the faster centroid approach for 
regions in the far field. This decision can be made on-the-fly in the geometry shader
based on the distance of the point coordinates to the camera. An
example of this is shown in Figure~\ref{Fig-LargeScale}.

Alternatively and/or in addition to this, a multi-resolution hierarchy,
for example an octree, can be constructed from the full-resolution 3D
position texture. A texel on the first coarser level would store the center
of mass as well as the averaged density of all tetrahedral elements
represented by the texels on the
highest level of resolution. The following coarser levels could then be
constructed from these using techniques like for example discussed in 
Fraedrich~et~al.~\cite{Fraedrich:2009:EMR:1638611.1639231}.
Again, regions close to the camera would be rendered via the
cell-projection approach using the original
resolution of the texture, whereas regions 
in the far field would be approximated using splatting techniques for
the coarser resolution textures.

\section{Conclusions}
\label{Sec-Conclusions}

We presented three GPU-accelerated rendering approaches for N-body dark matter
simulation data, based on a tetrahedral decomposition of the
computational domain that allows a physically more accurate
estimation of the mass density
between the tracer particles than previous methods.
They use the full phase-space information of the ensemble of dark
matter tracer particles and two of them minimize pre-processing time
(centroids and cell-projection) as well as  
data transfer between the CPU and GPU, by generating all
connectivity information as well as the derived quantities, like mass
density of the tetrahedral mesh elements, on the GPU. Thus these
approaches are particularly well suited for time-dependent data.
Their performance should benefit significantly from the 
increased number of cores expected for future generations of graphics hardware.

We compared these new methods to 
three standard rendering approaches for dark matter simulations: two  based on
constant and adaptive kernel sizes that estimate the local densities from 
the nearest-neighbors, as well as a Voronoi tessellation generated by
the simulations tracer particles. We showed that our approaches yield considerably
better image quality with less pre-processing times and graphics memory
requirements. The full tetrahedral cell-projection methods
clearly stands apart, however. Without artificial smoothing or density
estimates derived from averaging over the particle distribution,
features previously washed out, become clearly visible and give new
insight in the physical large-scale features of the {\sl cosmic web}, including
voids, filaments and halos.


\acknowledgments{
This work was supported support by the National Science
Foundation through award number AST-0808398 and the LDRD program at
the SLAC National Accelerator Laboratory as well as the Terman
fellowship at Stanford University.}

\bibliographystyle{abbrv}


\begin{thebibliography}{10}

\bibitem{2005ApJsloan}
K.~{Abazajian}, Z.~{Zheng}, I.~{Zehavi}, D.~H. {Weinberg}, J.~A. {Frieman},
  A.~A. {Berlind}, M.~R. {Blanton}, N.~A. {Bahcall}, J.~{Brinkmann}, D.~P.
  {Schneider}, and M.~{Tegmark}.
\newblock {Cosmology and the Halo Occupation Distribution from Small-Scale
  Galaxy Clustering in the Sloan Digital Sky Survey}.
\newblock {\em apj}, 625:613--620, June 2005.

\bibitem{2011AbelHahnKaehler}
T.~{Abel}, O.~{Hahn}, and R.~{Kaehler}.
\newblock {Tracing the Dark Matter Sheet in Phase Space}.
\newblock {\em arXiv}, 1111.3944v2, Nov. 2011.

\bibitem{1996MNRAS.279..693B}
F.~{Bernardeau} and R.~{van de Weygaert}.
\newblock {A New Method for Accurate Estimation of Velocity Field Statistics}.
\newblock {\em {MNRAS}}, 279:693, Mar. 1996.

\bibitem{2009MNRAS.398.1150B}
M.~{Boylan-Kolchin}, V.~{Springel}, S.~D.~M. {White}, A.~{Jenkins}, and
  G.~{Lemson}.
\newblock {Resolving cosmic structure formation with the Millennium-II
  Simulation}.
\newblock {\em mnras}, 398:1150--1164, Sept. 2009.

\bibitem{journals/cgf/ChaSI09}
D.~Cha, S.~Son, and I.~Ihm.
\newblock {GPU-Assisted High Quality Particle Rendering.}
\newblock {\em Comput. Graph. Forum}, 28(4):1247--1255, 2009.

\bibitem{1985ApJS...57..241E}
G.~{Efstathiou}, M.~{Davis}, S.~D.~M. {White}, and C.~S. {Frenk}.
\newblock {Numerical techniques for large cosmological N-body simulations}.
\newblock {\em apjs}, 57:241--260, Feb. 1985.

\bibitem{Espinha:2005:HHR:1114697.1115365}
R.~Espinha and W.~Celes.
\newblock {High-Quality Hardware-Based Ray-Casting Volume Rendering Using
  Partial Pre-Integration}.
\newblock In {\em Proceedings of the XVIII Brazilian Symposium on Computer
  Graphics and Image Processing}, pages 273--, Washington, DC, USA, 2005. IEEE
  Computer Society.

\bibitem{Fraedrich:2010:EHV:1907651.1907951}
R.~Fraedrich, S.~Auer, and R.~Westermann.
\newblock {Efficient High-Quality Volume Rendering of SPH Data}.
\newblock {\em IEEE Transactions on Visualization and Computer Graphics},
  16:1533--1540, November 2010.

\bibitem{Fraedrich:2009:EMR:1638611.1639231}
R.~Fraedrich, J.~Schneider, and R.~Westermann.
\newblock {Exploring the Millennium Run - Scalable Rendering of Large-Scale
  Cosmological Datasets}.
\newblock {\em IEEE Transactions on Visualization and Computer Graphics},
  15:1251--1258, November 2009.

\bibitem{Haroz2008}
S.~Haroz, K.-L. Ma, and K.~Heitmann.
\newblock Multiple uncertainties in time-variant cosmological particle data.
\newblock In {\em Proceedings of IEEE Pacific Visualization Symposium}, pages
  207--214. IEEE VGTC, March 2008.

\bibitem{Hilbert:2009}
S.~{Hilbert}, J.~{Hartlap}, S.~D.~M. {White}, and P.~{Schneider}.
\newblock {Ray-tracing through the Millennium Simulation: Born corrections and
  lens-lens coupling in cosmic shear and galaxy-galaxy lensing}.
\newblock {\em A\&A}, 499:31--43, May 2009.

\bibitem{Hopf:2003:HSS:1081432.1081501}
M.~Hopf and T.~Ertl.
\newblock {Hierarchical Splatting of Scattered Data}.
\newblock In {\em Proceedings of the 14th IEEE Visualization 2003 (VIS'03)},
  VIS '03, pages 57--, Washington, DC, USA, 2003. IEEE Computer Society.

\bibitem{Hopf:2004:HSS:1018014.1018059}
M.~Hopf, M.~Luttenberger, and T.~Ertl.
\newblock {Hierarchical Splatting of Scattered 4D Data}.
\newblock {\em IEEE Comput. Graph. Appl.}, 24:64--72, July 2004.

\bibitem{journals/procedia/JinKRGDR10}
Z.~Jin, M.~Krokos, M.~Rivi, C.~Gheller, K.~Dolag, and M.~Reinecke.
\newblock {High-performance Astrophysical Visualization using Splotch.}
\newblock {\em Procedia CS}, 1(1):1775--1784, 2010.

\bibitem{0004-637X-740-2-102}
A.~A. Klypin, S.~Trujillo-Gomez, and J.~Primack.
\newblock Dark matter halos in the standard cosmological model: Results from
  the bolshoi simulation.
\newblock {\em The Astrophysical Journal}, 740(2):102, 2011.

\bibitem{Kraus:2004:PTW:1032664.1034427}
M.~Kraus, W.~Qiao, and D.~S. Ebert.
\newblock Projecting tetrahedra without rendering artifacts.
\newblock In {\em Proceedings of the conference on Visualization '04}, VIS '04,
  pages 27--34, Washington, DC, USA, 2004. IEEE Computer Society.

\bibitem{CGF29-3:903-912:2010}
A.~Maximo, R.~Marroquim, and R.~Farias.
\newblock {Hardware-Assisted Projected Tetrahedra}.
\newblock {\em Computer Graphics Forum}, 29(3):903--912, 2010.

\bibitem{1988CoPhC..48...89M}
J.~J. {Monaghan}.
\newblock {An introduction to SPH}.
\newblock {\em Computer Physics Communications}, 48:89--96, Jan. 1988.

\bibitem{muigg-2011-gpg}
P.~Muigg, M.~Hadwiger, H.~Doleisch, and M.~E. Gr{\"o}ller.
\newblock {Interactive Volume Visualization of General Polyhedral Grids}.
\newblock {\em IEEE Transaction on Visualization and Computer Graphics},
  17(12):2115--2124, 12 2011.

\bibitem{Peebles1993}
P.~J.~E. {Peebles}.
\newblock {\em {Principles of Physical Cosmology}}.
\newblock 1993.

\bibitem{Popov-UCSC-SOE-11-17}
U.~Popov, K.~Heitmann, J.~Ahrens, S.~Habib, and A.~Pang.
\newblock {The Evolution of Multistreaming Events in the Formation of Large
  Scale Structures}.
\newblock {\em {Technical Report, UCSC}}, {}({UCSC-SOE-11-17}), 2011.

\bibitem{price-2007}
D.~J. Price.
\newblock {SPLASH: An Interactive Visualisation Tool for Smoothed Particle
  Hydrodynamics Simulations}, 2007.

\bibitem{voro++}
C.~Rycroft.
\newblock {The Voro++ Software Library}.
\newblock \url{http://math.lbl.gov/voro++/}.

\bibitem{Shandarin:2011jv}
S.~Shandarin, S.~Habib, and K.~Heitmann.
\newblock {The Cosmic Web, Multi-Stream Flows, and Tessellations}.
\newblock {\em arXiv}, 1111.2366, 2011.

\bibitem{Shirley:1990:PAD:99307.99322}
P.~Shirley and A.~Tuchman.
\newblock {A Polygonal Approximation to Direct Scalar Volume Rendering}.
\newblock In {\em Proceedings of the 1990 workshop on Volume visualization},
  VVS '90, pages 63--70, New York, NY, USA, 1990. ACM.

\bibitem{Springel05thecosmological}
V.~Springel.
\newblock {The Cosmological Simulation Code Gadget-2}.
\newblock {\em Monthly Notices of the Royal Astronomical Society}, 364, 2005.

\bibitem{szalay-2008}
T.~Szalay, V.~Springel, and G.~Lemson.
\newblock {GPU-Based Interactive Visualization of Billion Point Cosmological
  Simulations}, 2008.

\bibitem{Wang:2007a}
J.~{Wang} and S.~D.~M. {White}.
\newblock {Discreteness effects in simulations of hot/warm dark matter}.
\newblock {\em mnras}, 380:93--103, Sept. 2007.

\bibitem{Weiler:2001:HRI:601671.601702}
M.~Weiler and T.~Ertl.
\newblock Hardware-software-balanced resampling for the interactive
  visualization of unstructured grids.
\newblock In {\em Proceedings of the conference on Visualization '01}, VIS '01,
  pages 199--206, Washington, DC, USA, 2001. IEEE Computer Society.

\bibitem{Weiler:2003:HRC:1081432.1081488}
M.~Weiler, M.~Kraus, M.~Merz, and T.~Ertl.
\newblock {Hardware-Based Ray Casting for Tetrahedral Meshes}.
\newblock In {\em Proceedings of the 14th IEEE Visualization 2003 (VIS'03)},
  VIS '03, pages 44--, Washington, DC, USA, 2003. IEEE Computer Society.

\bibitem{westermann:2001:unstructured}
R.~Westermann.
\newblock The rendering of unstructured grids revisited.
\newblock In {\em EG/IEEE TCVG Symposium on Visualization (VisSym '01)}, 2001.

\bibitem{Wylie:2002:TPU:584110.584112}
B.~Wylie, K.~Moreland, L.~A. Fisk, and P.~Crossno.
\newblock Tetrahedral projection using vertex shaders.
\newblock In {\em Proceedings of the 2002 IEEE symposium on Volume
  visualization and graphics}, VVS '02, pages 7--12, Piscataway, NJ, USA, 2002.
  IEEE Press.

\bibitem{Zhou06interactivepoint-based}
Y.~Zhou and M.~Garland.
\newblock Interactive point-based rendering of higher-order tetrahedral data.
\newblock {\em IEEE Transactions on Visualization and Computer Graphics},
  12(5):2006, 2006.

\bibitem{Zhu:2005:ASF:1073204.1073298}
Y.~Zhu and R.~Bridson.
\newblock Animating sand as a fluid.
\newblock {\em ACM Trans. Graph.}, 24:965--972, July 2005.

\bibitem{1937ApJzwicky}
F.~{Zwicky}.
\newblock {On the Masses of Nebulae and of Clusters of Nebulae}.
\newblock {\em apj}, 86:217, Oct. 1937.

\end{thebibliography}
\end{document}